Softwire Hub and Spoke Deployment Framework
with Layer Two Tunneling Protocol Version 2 (L2TPv2)

Status of This Memo

   This document specifies an Internet standards track protocol for the
   Internet community, and requests discussion and suggestions for
   improvements.  Please refer to the current edition of the "Internet
   Official Protocol Standards" (STD 1) for the standardization state
   and status of this protocol.  Distribution of this memo is unlimited.

Abstract

   This document describes the framework of the Softwire "Hub and Spoke"
   solution with the Layer Two Tunneling Protocol version 2 (L2TPv2).
   The implementation details specified in this document should be
   followed to achieve interoperability among different vendor
   implementations.

Table of Contents

## 1. Introduction

   The Softwires Working Group has selected Layer Two Tunneling Protocol
   version 2 (L2TPv2) as the phase 1 protocol to be deployed in the
   Softwire "Hub and Spoke" solution space.  This document describes the
   framework for the L2TPv2 "Hub and Spoke" solution, and the
   implementation details specified in this document should be followed
   to achieve interoperability among different vendor implementations.

   In the "Hub and Spoke" solution space, a Softwire is established to
   provide the home network with IPv4 connectivity across an IPv6-only
   access network, or IPv6 connectivity across an IPv4-only access
   network.  When L2TPv2 is used in the Softwire context, the voluntary
   tunneling model applies.  The Softwire Initiator (SI) at the home
   network takes the role of the L2TP Access Concentrator (LAC) client
   (initiating both the L2TP tunnel/session and the PPP link) while the
   Softwire Concentrator (SC) at the ISP takes the role of the L2TP
   Network Server (LNS).  The terms voluntary tunneling and compulsory
   tunneling are defined in Section 1.1 of [RFC3193].  Since the L2TPv2
   compulsory tunneling model does not apply to Softwires, it SHOULD NOT
   be requested or honored.  This document identifies all the voluntary
   tunneling related L2TPv2 attributes that apply to Softwires and
   specifies the handling mechanism for such attributes in order to
   avoid ambiguities in implementations.  This document also identifies
   the set of L2TPv2 attributes specific to the compulsory tunneling
   model that does not apply to Softwires and specifies the mechanism to
   ignore or nullify their effect within the Softwire context.

   The SI and SC MUST follow the L2TPv2 operations described in
   [RFC2661] when performing Softwire establishment, teardown, and
   Operations, Administration, and Management (OAM).  With L2TPv2, a
   Softwire consists of an L2TPv2 Control Connection (also referred to
   as Control Channel), a single L2TPv2 Session, and the PPP link
   negotiated over the Session.  To establish the Softwire, the SI first
   initiates an L2TPv2 Control Channel to the SC, which accepts the
   request and terminates the Control Channel.  L2TPv2 supports an
   optional mutual Control Channel authentication that allows both SI
   and SC to validate each other's identity at the initial phase of
   hand-shaking before proceeding with Control Channel establishment.
   After the L2TPv2 Control Channel is established between the SI and
   SC, the SI initiates an L2TPv2 Session to the SC.  Then the PPP/IP
   link is negotiated over the L2TPv2 Session between the SI and SC.
   After the PPP/IP link is established, it acts as the Softwire between
   the SI and SC for tunneling IP traffic of one Address Family (AF)
   across the access network of another Address Family.

   During the life of the Softwire, both SI and SC send L2TPv2 keepalive
   HELLO messages to monitor the health of the Softwire and the peer
   L2TP Control Connection Endpoint (LCCE), and to potentially refresh
   the NAT/NAPT (Network Address Translation / Network Address Port
   Translation) entry at the CPE or at the other end of the access link.
   Optionally, Link Control Protocol (LCP) ECHO messages can be used as
   keepalives for the same purposes.  In the event of keepalive timeout
   or administrative shutdown of the Softwire, either the SI or the SC
   MAY tear down the Softwire by tearing down the L2TPv2 Control Channel
   and Session as specified in [RFC2661].

1.1.  Abbreviations

   AF     Address Family, IPv4 or IPv6.

   CPE    Customer Premises Equipment.

   LCCE   L2TP Control Connection Endpoint, an L2TP node that exists at
          either end of an L2TP Control Connection.  (See [RFC3931].)

   LNS    L2TP Network Server, a node that acts as one side of an L2TP
          tunnel (Control Connection) endpoint.  The LNS is the logical
          termination point of a PPP session that is being tunneled from
          the remote system by the peer LCCE.  (See [RFC2661].)

   SC     Softwire Concentrator, the node terminating the Softwire in
          the service provider network.  (See [RFC4925].)

   SI     Softwire Initiator, the node initiating the Softwire within
          the customer network.  (See [RFC4925].)

   SPH    Softwire Payload Header, the IP headers being carried within a
          Softwire.  (See [RFC4925].)

   STH    Softwire Transport Header, the outermost IP header of a
          Softwire.  (See [RFC4925].)

   SW     Softwire, a shared-state "tunnel" created between the SC and
          SI.  (See [RFC4925].)

1.2.  Requirements Language

   The key words "MUST", "MUST NOT", "REQUIRED", "SHALL", "SHALL NOT",
   "SHOULD", "SHOULD NOT", "RECOMMENDED", "MAY", and "OPTIONAL" in this
   document are to be interpreted as described in [RFC2119].

1.3. Considerations

   Some sections of this document contain considerations that are not
   required for interoperability and correct operation of Softwire
   implementations.  These sections' titles are marked beginning with
   "Considerations".

2.  Applicability of L2TPv2 for Softwire Requirements

   A list of Softwire "Hub and Spoke" requirements has been identified
   by the Softwire Problem Statement [RFC4925].  The following sub-
   sections describe how L2TPv2 fulfills each of them.

2.1.  Traditional Network Address Translation (NAT and NAPT)

   A "Hub and Spoke" Softwire must be able to traverse Network Address
   Translation (NAT) and Network Address Port Translation (NAPT, also
   referred to as Port Address Translation or PAT) devices [RFC3022] in
   case the scenario in question involves a non-upgradable, preexisting
   IPv4 home gateway performing NAT/NAPT or some carrier equipment at
   the other end of the access link performing NAT/NAPT.  The L2TPv2
   Softwire (i.e., Control Channel and Session) is capable of NAT/NAPT
   traversal since L2TPv2 can run over UDP.

   Since L2TPv2 does not detect NAT/NAPT along the path, L2TPv2 does not
   offer the option of disabling UDP.  The UDP encapsulation is present
   regardless of NAT/NAPT presence.  Both NAT/NAPT "autodetect" and UDP
   "bypass" are optional requirements in Section 2.3 of [RFC4925].

   As mentioned in Section 8.1 of [RFC2661] and Section 4 of [RFC3193],
   an L2TP Start-Control-Connection-Reply (SCCRP) responder (SC) can
   chose a different IP address and/or UDP port than those from the
   initiator's Start-Control-Connection-Request (SCCRQ) (SI).  This may
   or may not traverse a NAT/NAPT depending on the NAT/NAPT's Filtering
   Behavior (see Section 5 of [RFC4787]).  Specifically, any IP address
   and port combination will work with Endpoint-Independent Filtering,
   but changing the IP address and port will not work through Address-
   Dependent or Address-and-Port-Dependent Filtering.  Given this,
   responding from a different IP address and/or UDP port is NOT
   RECOMMENDED.

## [2.2](#). Scalability

   In the "Hub and Spoke" model, a carrier must be able to scale the
   solution to millions of Softwire Initiators by adding more hubs
   (i.e., Softwire Concentrators).  L2TPv2 is a widely deployed protocol
   in broadband services, and its scalability has been proven in
   multiple large-scale IPv4 Virtual Private Network deployments, which
   scale up to millions of subscribers each.

## [2.3](#). Routing

   There are no dynamic routing protocols between the SC and SI.  A
   default route from the SI to the SC is used.

## [2.4](#). Multicast

   Multicast protocols simply run transparently over L2TPv2 Softwires
   together with other regular IP traffic.

## [2.5](#). Authentication, Authorization, and Accounting (AAA)

   L2TPv2 supports optional mutual Control Channel authentication and
   leverages the optional mutual PPP per-session authentication.  L2TPv2
   is well integrated with AAA solutions (such as RADIUS) for both
   authentication and authorization.  Most L2TPv2 implementations
   available in the market support the logging of authentication and
   authorization events.

   L2TPv2 integration with RADIUS accounting (RADIUS Accounting
   extension for tunnel [[RFC2867](#)]) allows the collection and reporting
   of L2TPv2 Softwire usage statistics.

## [2.6](#). Privacy, Integrity, and Replay Protection

   Since L2TPv2 runs over IP/UDP in the Softwire context, IPsec
   Encapsulating Security Payload (ESP) can be used in conjunction to
   provide per-packet authentication, integrity, replay protection, and
   confidentiality for both L2TPv2 control and data traffic [[RFC3193](#)]
   and [[RFC3948](#)].

   For Softwire deployments in which full payload security is not
   required, the L2TPv2 built-in Control Channel authentication and the
   inherited PPP authentication and PPP Encryption Control Protocol can
   be considered.

## 2.7. Operations and Management

   L2TPv2 supports an optional in-band keepalive mechanism that injects
   HELLO control messages after a specified period of time has elapsed
   since the last data or control message was received on a tunnel (see
   [Section 5.5 of [RFC2661]](#)).  If the HELLO control message is not
   reliably delivered, then the Control Channel and its Session will be
   torn down.  In the Softwire context, the L2TPv2 keepalive is used to
   monitor the connectivity status between the SI and SC and/or as a
   refresh mechanism for any NAT/NAPT translation entry along the access
   link.

   LCP ECHO offers a similar mechanism to monitor the connectivity
   status, as described in [[RFC1661](#)].  Softwire implementations SHOULD
   use L2TPv2 Hello keepalives, and in addition MAY use PPP LCP Echo
   messages to ensure Dead End Detection and/or to refresh NAT/NAPT
   translation entries.  The combination of these two mechanisms can be
   used as an optimization.

   The L2TPv2 MIB [[RFC3371](#)] supports the complete suite of management
   operations such as configuration of Control Channel and Session,
   polling of Control Channel and Session status and their traffic
   statistics and notifications of Control Channel and Session UP/DOWN
   events.

## 2.8. Encapsulations

   L2TPv2 supports the following encapsulations:

   o  IPv6/PPP/L2TPv2/UDP/IPv4

   o  IPv4/PPP/L2TPv2/UDP/IPv6

   o  IPv4/PPP/L2TPv2/UDP/IPv4

   o  IPv6/PPP/L2TPv2/UDP/IPv6

   Note that UDP bypass is not supported by L2TPv2 since L2TPv2 does not
   support "autodetect" of NAT/NAPT.

## 3.  Deployment Scenarios

   For the "Hub and Spoke" problem space, four scenarios have been
   identified.  In each of these four scenarios, different home
   equipment plays the role of the Softwire Initiator.  This section
   elaborates each scenario with L2TPv2 as the Softwire protocol and

   other possible protocols involved to complete the solution.  This
   section examines the four scenarios for both IPv6-over-IPv4
   (Section 3.1) and IPv4-over-IPv6 (Section 3.2) encapsulations.

3.1.  IPv6-over-IPv4 Softwires with L2TPv2

   The following sub-sections cover IPv6 connectivity (SPH) across an
   IPv4-only access network (STH) using a Softwire.

3.1.1.  Host CPE as Softwire Initiator

   The Softwire Initiator (SI) is the host CPE (directly connected to a
   modem), which is dual-stack.  There is no other gateway device.  The
   IPv4 traffic SHOULD NOT traverse the Softwire.  See Figure 1.

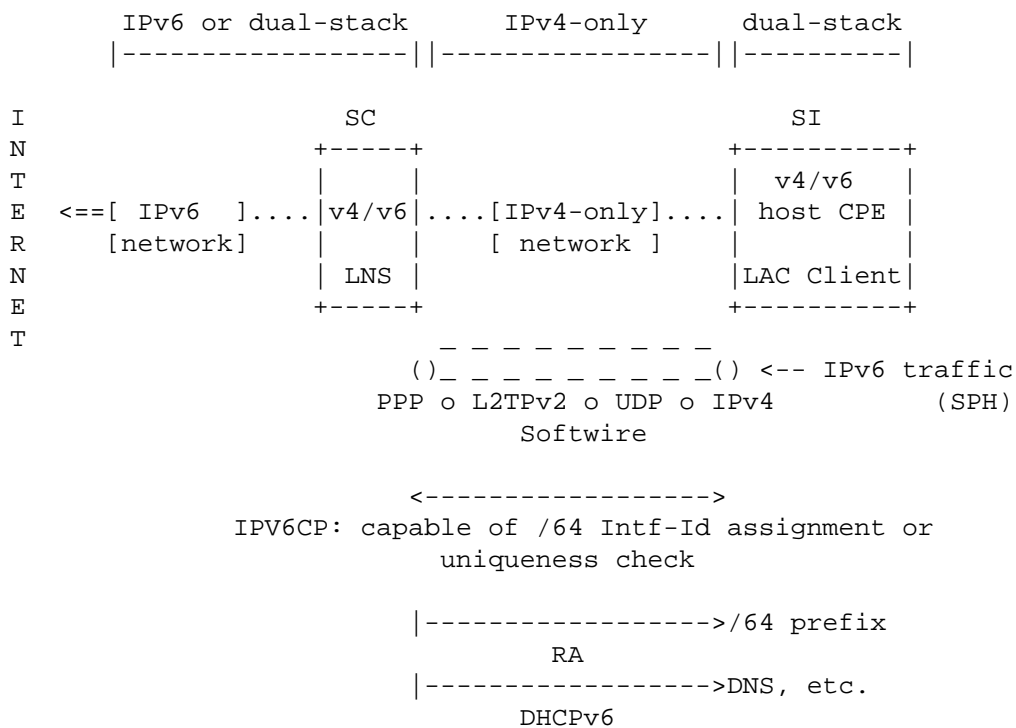

              Figure 1: Host CPE as Softwire Initiator

   In this scenario, after the L2TPv2 Control Channel and Session
   establishment and PPP LCP negotiation (and optionally PPP
   Authentication) are successful, the IPv6 Control Protocol (IPV6CP)
   negotiates IPv6-over-PPP, which also provides the capability for the
   ISP to assign the 64-bit Interface-Identifier to the host CPE or
   perform uniqueness validation for the two interface identifiers at
   the two PPP ends [RFC5072].  After IPv6-over-PPP is up, IPv6

   Stateless Address Autoconfiguration / Neighbor Discovery runs over
   the IPv6-over-PPP link, and the LNS can inform the host CPE of a
   prefix to use for stateless address autoconfiguration through a
   Router Advertisement (RA) while other non-address configuration
   options (such as DNS [RFC3646] or other servers' addresses that might
   be available) can be conveyed to the host CPE via DHCPv6.

3.1.2.  Router CPE as Softwire Initiator

   The Softwire Initiator (SI) is the router CPE, which is a dual-stack
   device.  The IPv4 traffic SHOULD NOT traverse the Softwire.  See
   Figure 2.

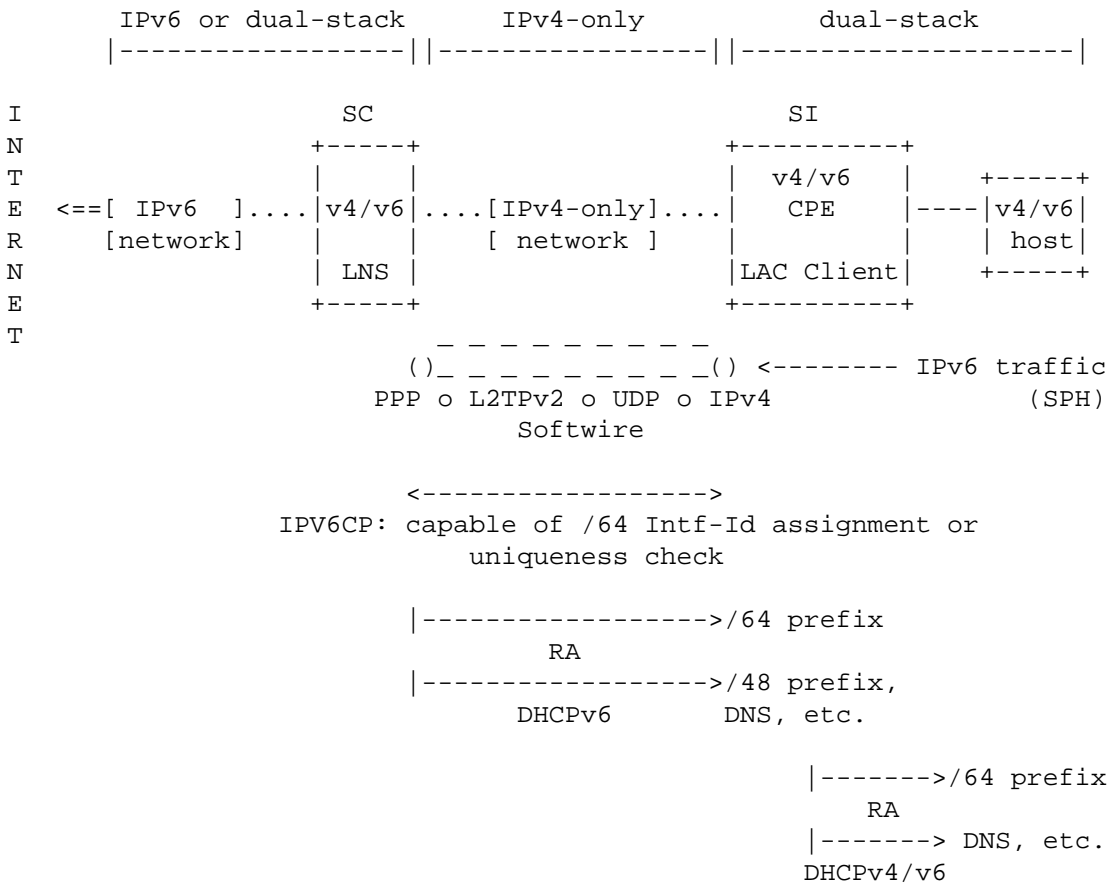

                Figure 2: Router CPE as Softwire Initiator

   In this scenario, after the L2TPv2 Control Channel and Session
   establishment and PPP LCP negotiation (and optionally PPP
   Authentication) are successful, IPV6CP negotiates IPv6-over-PPP,
   which also provides the capability for the ISP to assign the 64-bit

Interface-Identifier to the router CPE or perform uniqueness
validation for the two interface identifiers at the two PPP ends
[RFC5072].  After IPv6-over-PPP is up, IPv6 Stateless Address
Autoconfiguration / Neighbor Discovery runs over the IPv6-over-PPP
link, and the LNS can inform the router CPE of a prefix to use for
stateless address autoconfiguration through a Router Advertisement
(RA).  DHCPv6 can be used to perform IPv6 Prefix Delegation (e.g.,
delegating a prefix to be used within the home network [RFC3633]) and
convey other non-address configuration options (such as DNS
[RFC3646]) to the router CPE.

3.1.3.  Host behind CPE as Softwire Initiator

   The CPE is IPv4-only.  The Softwire Initiator (SI) is a dual-stack
   host (behind the IPv4-only CPE), which acts as an IPv6 host CPE.  The
   IPv4 traffic SHOULD NOT traverse the Softwire.  See Figure 3.

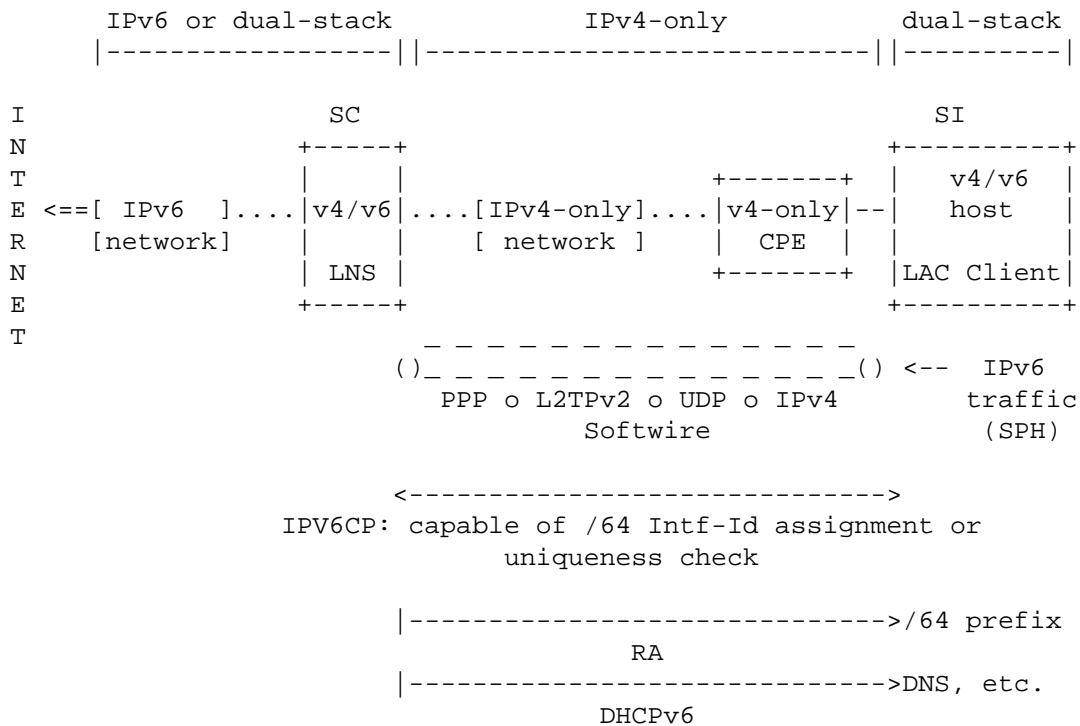

              Figure 3: Host behind CPE as Softwire Initiator

   In this scenario, after the L2TPv2 Control Channel and Session
   establishment and PPP LCP negotiation (and optionally PPP
   Authentication) are successful, IPV6CP negotiates IPv6-over-PPP,
   which also provides the capability for the ISP to assign the 64-bit

   Interface-Identifier to the host or perform uniqueness validation for
   the two interface identifiers at the two PPP ends [RFC5072].  After
   IPv6-over-PPP is up, IPv6 Stateless Address Autoconfiguration /
   Neighbor Discovery runs over the IPv6-over-PPP link, and the LNS can
   inform the host of a prefix to use for stateless address
   autoconfiguration through a Router Advertisement (RA) while other
   non-address configuration options (such as DNS [RFC3646]) can be
   conveyed to the host via DHCPv6.

3.1.4.  Router behind CPE as Softwire Initiator

   The CPE is IPv4-only.  The Softwire Initiator (SI) is a dual-stack
   device (behind the IPv4-only CPE) acting as an IPv6 CPE router inside
   the home network.  The IPv4 traffic SHOULD NOT traverse the Softwire.
   See Figure 4.

```
            IPv6 or dual-stack       IPv4-only         dual-stack
           |-----------------||-------------------------||------------|

  I                         SC                                   SI
  N                       +-----+                           +---------+
  T                       |     |           +-------+       |  v4/v6  |
  E   <==[ IPv6   ]....|v4/v6|..[IPv4-only]..|v4-only|---|  router  |
  R     [network]       |     |  [ network ]  |  CPE  |    |         |
  N                     | LNS |                +-------+    |LAC Client|
  E                     +-----+                             +---------+
  T                                                          |
                                                     --------+-----+
                                                             |v4/v6|
                                                             | host|
                                                             +-----+
                            _ _ _ _ _ _ _ _ _ _ _
                        ()_ _ _ _ _ _ _ _ _ _ _()  <-- IPv6
                         PPP o L2TPv2 o UDP o IPv4   traffic
                                 Softwire             (SPH)

                         <--------------------------->
                         IPV6CP: capable of /64 Intf-Id assignment or
                                      uniqueness check

                              |-------------------------->/64 prefix
                                           RA
                              |-------------------------->/48 prefix,
                                         DHCPv6            DNS, etc.

                                                    |----> /64
                                                      RA   prefix
                                                    |----> DNS,
                                                     DHCPv6 etc.
```

             Figure 4: Router behind CPE as Softwire Initiator

   In this scenario, after the L2TPv2 Control Channel and Session
   establishment and PPP LCP negotiation (and optionally PPP
   Authentication) are successful, IPV6CP negotiates IPv6-over-PPP,
   which also provides the capability for the ISP to assign the 64-bit
   Interface-Identifier to the v4/v6 router or perform uniqueness
   validation for the two interface identifiers at the two PPP ends
   [RFC5072].  After IPv6-over-PPP is up, IPv6 Stateless Address
   Autoconfiguration / Neighbor Discovery runs over the IPv6-over-PPP
   link, and the LNS can inform the v4/v6 router of a prefix to use for
   stateless address autoconfiguration through a Router Advertisement

   (RA).  DHCPv6 can be used to perform IPv6 Prefix Delegation (e.g.,
   delegating a prefix to be used within the home network [RFC3633]) and
   convey other non-address configuration options (such as DNS
   [RFC3646]) to the v4/v6 router.

3.2.  IPv4-over-IPv6 Softwires with L2TPv2

   The following sub-sections cover IPv4 connectivity (SPH) across an
   IPv6-only access network (STH) using a Softwire.

3.2.1.  Host CPE as Softwire Initiator

   The Softwire Initiator (SI) is the host CPE (directly connected to a
   modem), which is dual-stack.  There is no other gateway device.  The
   IPv6 traffic SHOULD NOT traverse the Softwire.  See Figure 5.

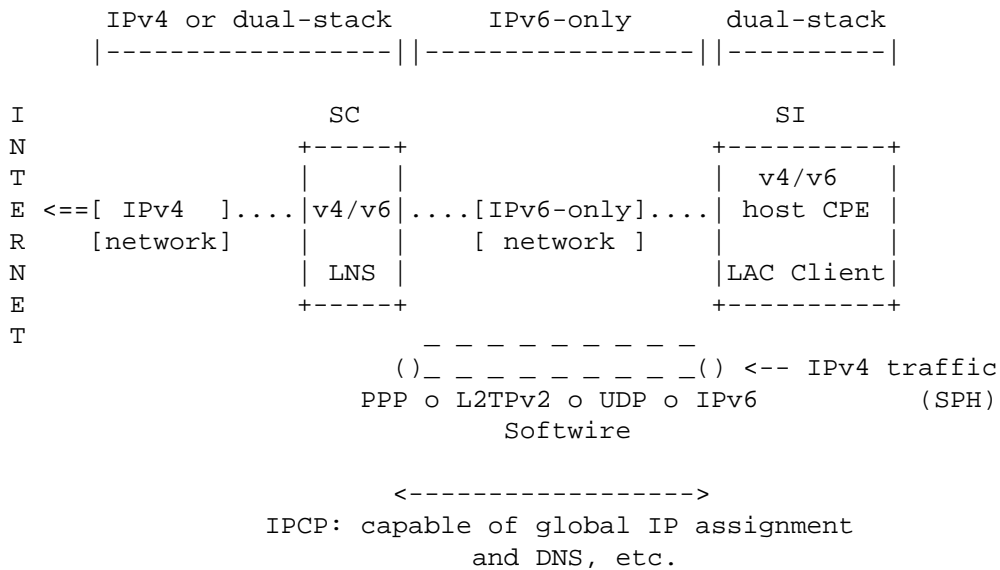

                Figure 5: Host CPE as Softwire Initiator

   In this scenario, after the L2TPv2 Control Channel and Session
   establishment and PPP LCP negotiation (and optionally PPP
   Authentication) are successful, the IP Control Protocol (IPCP)
   negotiates IPv4-over-PPP, which also provides the capability for the
   ISP to assign a global IPv4 address to the host CPE.  A global IPv4
   address can also be assigned via DHCP.  Other configuration options
   (such as DNS) can be conveyed to the host CPE via IPCP [RFC1877] or
   DHCP [RFC2132].

3.2.2.  Router CPE as Softwire Initiator

   The Softwire Initiator (SI) is the router CPE, which is a dual-stack
   device.  The IPv6 traffic SHOULD NOT traverse the Softwire.  See
   Figure 6.

```
         IPv4 or dual-stack      IPv6-only          dual-stack Home
         |------------------||-----------------||------------------|

    I                        SC                      SI
    N                      +-----+                +----------+
    T                      |     |                |  v4/v6   |   +-----+
    E  <==[  IPv4  ]....|v4/v6|....[IPv6-only]....|   CPE    |--|v4/v6|
    R     [network]    |     |     [ network ]   |          |  | host|
    N                      | LNS |                |LAC Client|   +-----+
    E                      +-----+                +----------+
    T                                 _ _ _ _ _ _ _ _
                                   ()_ _ _ _ _ _ _ _() <--------- IPv4 traffic
                                    PPP o L2TPv2 o UDP o IPv6              (SPH)
                                              Softwire

                              <------------------>
                           IPCP: capable of global IP assignment
                                    and DNS, etc.

                              |------------------>
                           DHCPv4: prefix, mask, PD

                                                              private/
                                                   |------> global
                                                     DHCP   IP, DNS,
                                                            etc.
```

              Figure 6: Router CPE as Softwire Initiator

   In this scenario, after the L2TPv2 Control Channel and Session
   establishment and PPP LCP negotiation (and optionally PPP
   Authentication) are successful, IPCP negotiates IPv4-over-PPP, which
   also provides the capability for the ISP to assign a global IPv4
   address to the router CPE.  A global IPv4 address can also be
   assigned via DHCP.  Other configuration options (such as DNS) can be
   conveyed to the router CPE via IPCP [RFC1877] or DHCP [RFC2132].  For
   IPv4 Prefix Delegation for the home network, DHCP [SUBNET-ALL] can be
   used.

3.2.3.  Host behind CPE as Softwire Initiator

   The CPE is IPv6-only.  The Softwire Initiator (SI) is a dual-stack
   host (behind the IPv6 CPE), which acts as an IPv4 host CPE.  The IPv6
   traffic SHOULD NOT traverse the Softwire.  See Figure 7.

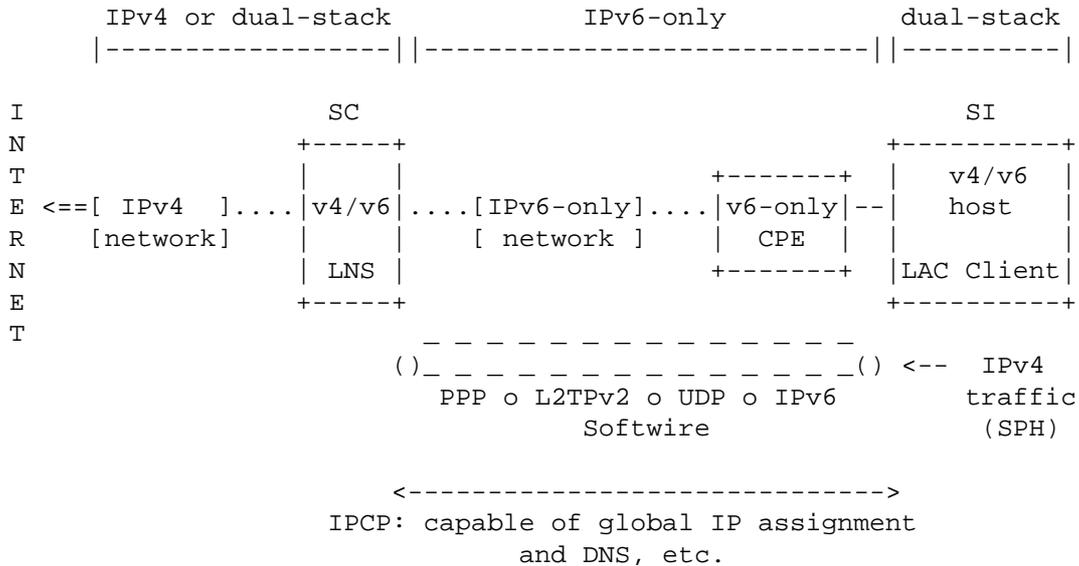

```
         IPv4 or dual-stack         IPv6-only              dual-stack
         |-----------------||---------------------------||----------|

     I                SC                                       SI
     N                +-----+                              +----------+
     T                |     |                  +-------+   | v4/v6    |
     E  <==[ IPv4  ]....|v4/v6|....[IPv6-only]....|v6-only|--| host     |
     R     [network]  |     |    [ network ]   | CPE   |   |          |
     N                | LNS |                  +-------+   |LAC Client|
     E                +-----+                              +----------+
     T                         _ _ _ _ _ _ _ _ _ _ _ _
                              ()_ _ _ _ _ _ _ _ _ _ _ _() <-- IPv4
                              PPP o L2TPv2 o UDP o IPv6      traffic
                                       Softwire               (SPH)

                         <------------------------------>
                         IPCP: capable of global IP assignment
                                   and DNS, etc.
```

          Figure 7: Host behind CPE as Softwire Initiator

   In this scenario, after the L2TPv2 Control Channel and Session
   establishment and PPP LCP negotiation (and optionally PPP
   Authentication) are successful, IPCP negotiates IPv4-over-PPP, which
   also provides the capability for the ISP to assign a global IPv4
   address to the host.  A global IPv4 address can also be assigned via
   DHCP.  Other configuration options (such as DNS) can be conveyed to
   the host CPE via IPCP [RFC1877] or DHCP [RFC2132].

3.2.4.  Router behind CPE as Softwire Initiator

   The CPE is IPv6-only.  The Softwire Initiator (SI) is a dual-stack
   device (behind the IPv6-only CPE) acting as an IPv4 CPE router inside
   the home network.  The IPv6 traffic SHOULD NOT traverse the Softwire.
   See Figure 8.

```
         IPv4 or dual-stack      IPv6-only         dual-stack
         |-----------------||------------------------||------------|

   I                        SC                            SI
   N                       +-----+                     +----------+
   T                       |     |          +-------+  |  v4/v6   |
   E  <==[ IPv4  ]....|v4/v6|..[IPv6-only]..|v6-only|---| router   |
   R    [network]    |     |  [ network ]   | CPE  |  |          |
   N                 | LNS |                +-------+ |LAC Client|
   E                 +-----+                          +----------+
   T                                                      |
                                                      --------+-----+
                                                      |v4/v6|
                                                      | host|
                                        _ _ _ _ _ _ _ _ _ _ _   +-----+
                                    ()_ _ _ _ _ _ _ _ _ _ _ _() <--- IPv4
                                     PPP o L2TPv2 o UDP o IPv4     traffic
                                              Softwire                (SPH)

                            <---------------------------->
                       IPCP: assigns global IP address and DNS, etc.

                           |---------------------------->
                            DHCPv4: prefix, mask, PD

                                                          private/
                                                   |----> global
                                                     DHCP  IP, DNS,
                                                           etc.
```

Figure 8: Router behind CPE as Softwire Initiator

In this scenario, after the L2TPv2 Control Channel and Session
establishment and PPP LCP negotiation (and optionally PPP
Authentication) are successful, IPCP negotiates IPv4-over-PPP, which
also provides the capability for the ISP to assign a global IPv4
address to the v4/v6 router.  A global IPv4 address can also be
assigned via DHCP.  Other configuration options (such as DNS) can be
conveyed to the v4/v6 router via IPCP [RFC1877] or DHCP [RFC2132].
For IPv4 Prefix Delegation for the home network, DHCP [SUBNET-ALL]
can be used.

4.  References to Standardization Documents

   This section lists and groups documents from the Internet
   standardization describing technologies used to design the framework
   of the Softwire "Hub and Spoke" solution.  This emphasizes the
   motivation of Softwire to reuse as many existing standards as

   possible.  This list contains both Standards Track (Proposed
   Standard, Draft Standard, and Standard) and Informational documents.
   The list of documents and their status should only be only used for
   description purposes.

4.1.  L2TPv2

   RFC 2661   "Layer Two Tunneling Protocol 'L2TP'" [RFC2661].

              *  For both IPv4 and IPv6 payloads (SPH), support is
                 complete.

              *  For both IPv4 and IPv6 transports (STH), support is
                 complete.

4.2.  Securing the Softwire Transport

   RFC 3193   "Securing L2TP using IPsec" [RFC3193].

   RFC 3948   "UDP Encapsulation of IPsec ESP Packets" [RFC3948].

              *   IPsec supports both IPv4 and IPv6 transports.

4.3.  Authentication, Authorization, and Accounting

   RFC 2865   "Remote Authentication Dial In User Service (RADIUS)"
              [RFC2865].

              *  Updated by [RFC2868], [RFC3575], and [RFC5080].

   RFC 2867   "RADIUS Accounting Modifications for Tunnel Protocol
              Support" [RFC2867].

   RFC 2868   "RADIUS Attributes for Tunnel Protocol Support" [RFC2868].

   RFC 3162   "RADIUS and IPv6" [RFC3162].

4.4.  MIB

   RFC 1471   "The Definitions of Managed Objects for the Link Control
              Protocol of the Point-to-Point Protocol" [RFC1471].

   RFC 1473   "The Definitions of Managed Objects for the IP Network
              Control Protocol of the Point-to-Point Protocol"
              [RFC1473].

   RFC 3371    "Layer Two Tunneling Protocol "L2TP" Management
               Information Base" [RFC3371].

   RFC 4087    "IP Tunnel MIB" [RFC4087].

               *  Both IPv4 and IPv6 transports are supported.

## 4.5.  Softwire Payload Related

### 4.5.1.  For IPv6 Payloads

   RFC 4861    "Neighbor Discovery for IP version 6 (IPv6)" [RFC4861].

   RFC 4862    "IPv6 Stateless Address Autoconfiguration" [RFC4862].

   RFC 5072    "IP Version 6 over PPP" [RFC5072].

   RFC 3315    "Dynamic Host Configuration Protocol for IPv6 (DHCPv6)"
               [RFC3315].

   RFC 3633    "IPv6 Prefix Options for Dynamic Host Configuration
               Protocol (DHCP) version 6" [RFC3633].

   RFC 3646    "DNS Configuration options for Dynamic Host Configuration
               Protocol for IPv6 (DHCPv6)" [RFC3646].

   RFC 3736    "Stateless Dynamic Host Configuration Protocol (DHCP)
               Service for IPv6" [RFC3736].

### 4.5.2.  For IPv4 Payloads

   RFC 1332    "The PPP Internet Protocol Control Protocol (IPCP)"
               [RFC1332].

   RFC 1661    "The Point-to-Point Protocol (PPP)" [RFC1661].

   RFC 1877    "PPP Internet Protocol Control Protocol Extensions for
               Name Server Addresses" [RFC1877].

   RFC 2131    "Dynamic Host Configuration Protocol" [RFC2131].

   RFC 2132    "DHCP Options and BOOTP Vendor Extensions" [RFC2132].

   DHCP Subnet Allocation  "Subnet Allocation Option".

               *  Work in progress, see [SUBNET-ALL].

## 5. Softwire Establishment

A Softwire is established in three distinct steps, potentially
preceded by an optional IPsec-related step 0 (see Figure 9).  First,
an L2TPv2 tunnel with a single session is established from the SI to
the SC.  Second, a PPP session is established over the L2TPv2 session
and the SI obtains an address.  Third, the SI optionally gets other
information through DHCP such as a delegated prefix and DNS servers.

```
   SC                                   SI
    |                                    |
    |<-------------IKEv1---------------->| Step 0
    |                                    | IPsec SA establishment
    |                                    | (optional)
    |                                    |
    |<-------------L2TPv2--------------->| Step 1
    |                                    | L2TPv2 Tunnel establishment
    |                                    |
    |<--------------PPP----------------->| Step 2
    |<-----Endpoint Configuration------->| PPP and Endpoint
    |                                    | configuration
    |                                    |
    |<------Router Configuration-------->| Step 3
    |                                    | Additional configuration
    |                                    | (optional)
```

         Figure 9: Steps for the Establishment of a Softwire

Figure 10 depicts details of each of these steps required to
establish a Softwire.

```
         SC                                          SI
          |                                           |
          |                                           |
          |                                           | Step 0
          |<------------IKEv1-------------->|         | = IKEv1 (Optional)
          |                                           |
          |                                           |
          |                                           | Step 1
          |<------------SCCRQ---------------|         -
          |-------------SCCRP-------------->|         |
          |<------------SCCCN---------------|         |
          |<------------ICRQ----------------|         | L2TPv2
          |-------------ICRP--------------->|         |
          |<------------ICCN----------------|         -
          |                                           |
          |                                           | Step 2
          |<-----Configuration-Request------|         -
          |------Configuration-Request----->|         | PPP
          |--------Configuration-Ack------->|         | LCP
          |<-------Configuration-Ack--------|         -
          |                                           |
          |-----------Challenge------------>|         - PPP Authentication
          |<----------Response--------------|         | (Optional - CHAP)
          |------------Success------------->|         -
          |                                           |
          |<-----Configuration-Request------|         -
          |------Configuration-Request----->|         | PPP NCP
          |--------Configuration-Ack------->|         | (IPV6CP or IPCP)
          |<-------Configuration-Ack--------|         -
          |                                           |
          |<------Router-Solicitation-------|         - Neighbor Discovery
          |-------Router-Advertisement----->|         | (IPv6 only)
          |                                           -
          |                                           |
          |                                           | Step3
          |                                           | DHCP (Optional)
          |<-----------SOLICIT--------------|         -
          |-----------ADVERTISE------------>|         | DHCPv6
          |<---------- REQUEST--------------|         | (IPv6 SW, Optional)
          |------------REPLY--------------->|         -
          |                                           | or
          |<---------DHCPDISCOVER-----------|         -
          |-----------DHCPOFFER------------>|         | DHCPv4
          |<---------DHCPREQUEST------------|         | (IPv4 SW, Optional)
          |-----------DHCPACK-------------->|         -
```

   Figure 10: Detailed Steps in the Establishment of a Softwire

   The IPsec-related negotiations in step 0 are optional.  The L2TPv2
   negotiations in step 1 are described in Section 5.1.  The PPP Network
   Control Protocol (NCP) negotiations in step 2 use IPV6CP for IPv6-
   over-IPv4 Softwires, and IPCP for IPv4-over-IPv6 Softwires (see
   Section 5.2.4).  The optional DHCP negotiations in step 3 use DHCPv6
   for IPv6-over-IPv4 Softwires, and DHCPv4 for IPv4-over-IPv6 Softwires
   (see Section 5.4).  Additionally, for IPv6-over-IPv4 Softwires, the
   DHCPv6 exchange for non-address configuration (such as DNS) can use
   Stateless DHCPv6, the two-message exchange with Information-Request
   and Reply messages (see Section 1.2 of [RFC3315] and [RFC3736]).

5.1.  L2TPv2 Tunnel Setup

   L2TPv2 [RFC2661] was originally designed to provide private network
   access to end users connected to a public network.  In the L2TPv2
   incoming call model, the end user makes a connection to an L2TP
   Access Concentrator (LAC).  The LAC then initiates an L2TPv2 tunnel
   to an L2TP Network Server (LNS).  The LNS then transfers end-user
   traffic between the L2TPv2 tunnel and the private network.

   In the Softwire "Hub and Spoke" model, the Softwire Initiator (SI)
   assumes the role of the LAC Client and the Softwire Concentrator (SC)
   assumes the role of the LNS.

   In the Softwire model, an L2TPv2 packet MUST be carried over UDP.
   The underlying version of the IP protocol may be IPv4 or IPv6,
   depending on the Softwire scenario.

   In the following sections, the term "Tunnel" follows the definition
   from Section 1.2 of [RFC2661], namely: "The Tunnel consists of a
   Control Connection and zero or more L2TP Sessions".

5.1.1.  Tunnel Establishment

   Figure 11 describes the messages exchanged and Attribute Value Pairs
   (AVPs) used to establish a tunnel between an SI (LAC) and an SC
   (LNS).  The messages and AVPs described here are only a subset of
   those defined in [RFC2661].  This is because Softwires use only a
   subset of the L2TPv2 functionality.  The subset of L2TP Control
   Connection Management AVPs that is applicable to Softwires is grouped
   into Required AVPs and Optional AVPs on a per-control-message basis
   (see Figure 11).  For each control message, Required AVPs include all
   the "MUST be present" AVPs from [RFC2661] for that control message,
   and Optional AVPs include the "MAY be present" AVPs from [RFC2661]
   that are used in the Softwire context on that control message.  Note
   that in the Softwire environment, the SI always initiates the tunnel.
   L2TPv2 AVPs SHOULD NOT be hidden.

```
                     SC                          SI
                      |<--------SCCRQ---------|
                        Required AVPs:
                            Message Type
                            Protocol Version
                            Host Name
                            Framing Capabilities
                            Assigned Tunnel ID
                        Optional AVPs:
                            Receive Window Size
                            Challenge
                            Firmware Revision
                            Vendor Name

                      |---------SCCRP-------->|
                        Required AVPs:
                            Message Type
                            Protocol Version
                            Framing Capabilities
                            Host Name
                            Assigned Tunnel ID
                        Optional AVPs:
                            Firmware Revision
                            Vendor Name
                            Receive Window Size
                            Challenge
                            Challenge Response

                      |<--------SCCCN---------|
                        Required AVPs:
                            Message Type
                        Optional AVPs:
                            Challenge Response
```

                Figure 11: Control Connection Establishment

   In L2TPv2, generally, the tunnel between an LAC and LNS may carry the
   data of multiple users.  Each of these users is represented by an
   L2TPv2 session within the tunnel.  In the Softwire environment, the
   tunnel carries the information of a single user.  Consequently, there
   is only one L2TPv2 session per tunnel.  Figure 12 describes the
   messages exchanged and the AVPs used to establish a session between
   an SI (LAC) and an SC (LNS).  The messages and AVPs described here
   are only a subset of those defined in [RFC2661].  This is because
   Softwires use only a subset of the L2TPv2 functionality.  The subset
   of L2TP Call Management (i.e., Session Management) AVPs that is
   applicable to Softwires is grouped into Required AVPs and Optional
   AVPs on a per-control-message basis (see Figure 12).  For each

   control message, Required AVPs include all the "MUST be present" AVPs
   from [RFC2661] for that control message, and Optional AVPs include
   the "MAY be present" AVPs from [RFC2661] that are used in the
   Softwire context on that control message.  Note that in the Softwire
   environment, the SI always initiates the session.  An L2TPv2 session
   setup for a Softwire uses only the incoming call model.  No outgoing
   or analog calls (sessions) are permitted.  L2TPv2 AVPs SHOULD NOT be
   hidden.

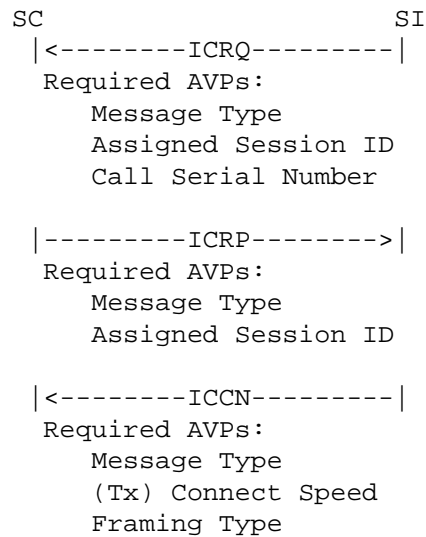

```
                   SC                        SI
                    |<--------ICRQ---------|
                      Required AVPs:
                         Message Type
                         Assigned Session ID
                         Call Serial Number

                    |---------ICRP-------->|
                      Required AVPs:
                         Message Type
                         Assigned Session ID

                    |<--------ICCN---------|
                      Required AVPs:
                         Message Type
                         (Tx) Connect Speed
                         Framing Type
```

                      Figure 12: Session Establishment

   The following sub-sections (5.1.1.1 through 5.1.1.3) describe in more
   detail the Control Connection and Session establishment AVPs (see
   message flows in Figures 11 and 12, respectively) that are required,
   optional and not relevant for the L2TPv2 Tunnel establishment of a
   Softwire.  Specific L2TPv2 protocol messages and flows that are not
   explicitly described in these sections are handled as defined in
   [RFC2661].

   The mechanism for hiding AVP Attribute values is used, as described
   in Section 4.3 of [RFC2661], to hide sensitive control message data
   such as usernames, user passwords, or IDs, instead of sending the AVP
   contents in the clear.  Since AVPs used in L2TP messages for the
   Softwire establishment do not transport such sensitive data, L2TPv2
   AVPs SHOULD NOT be hidden.

5.1.1.1. AVPs Required for Softwires

   This section prescribes specific values for AVPs that are required
   (by [RFC2661]) to be present in one or more of the messages used for
   the Softwire establishment, as they are used in the Softwire context.
   It combines all the Required AVPs from all the control messages in
   Section 5.1.1, and provides Softwire-specific use guidance.

   Host Name AVP

      This AVP is required in SCCRQ and SCCRP messages.  This AVP MAY be
      used to authenticate users, in which case it would contain a user
      identification.  If this AVP is not used to authenticate users, it
      may be used for logging purposes.

   Framing Capabilities AVP

      Both the synchronous (S) and asynchronous (A) bits SHOULD be set
      to 1.  This AVP SHOULD be ignored by the receiver.

   Framing Type AVP

      The synchronous bit SHOULD be set to 1 and the asynchronous bit to
      0.  This AVP SHOULD be ignored by the receiver.

   (Tx) Connect Speed AVP

      (Tx) Connect Speed is a required AVP but is not meaningful in the
      Softwire context.  Its value SHOULD be set to 0 and ignored by the
      receiver.

   Message Type AVP, Protocol Version AVP, Assigned Tunnel ID AVP, Call
   Serial Number AVP, and Assigned Session ID AVP

      As defined in [RFC2661].

5.1.1.2. AVPs Optional for Softwires

   This section prescribes specific values for AVPs that are Optional
   (not required by [RFC2661]) but used in the Softwire context.  It
   combines all the Optional AVPs from all the control messages in
   Section 5.1.1, and provides Softwire-specific use guidance.

   Challenge AVP and Challenge Response AVP

      These AVPs are not required, but are necessary to implement tunnel
      authentication.  Since tunnel authentication happens at the
      beginning of L2TPv2 tunnel creation, it can be helpful in
      preventing denial-of-service (DoS) attacks.  See Section 5.1.1 of
      [RFC2661].

      The usage of these AVPs in L2TP messages is OPTIONAL, but SHOULD
      be implemented in the SC.

   Receive Window Size AVP, Firmware Revision AVP, and Vendor Name AVP

      As defined in [RFC2661].

5.1.1.3.  AVPs Not Relevant for Softwires

   L2TPv2 specifies numerous AVPs that, while allowed for a given
   message, are irrelevant to Softwires.  They can be irrelevant to
   Softwires because they do not apply to the Softwire establishment
   flow (e.g., they are only used in the Outgoing Call establishment
   message exchange, while Softwires only use the Incoming Call message
   flow), or because they are Optional AVPs that are not used.  L2TPv2
   AVPs that are relevant to Softwires were covered in Sections 5.1.1,
   5.1.1.1, and 5.1.1.2.  Softwire implementations SHOULD NOT send AVPs
   that are not relevant to Softwires.  However, they SHOULD ignore them
   when they are received.  This will simplify the creation of Softwire
   applications that build upon existing L2TPv2 implementations.

5.1.2.  Tunnel Maintenance

   Periodically, the SI/SC MUST transmit a message to the peer to detect
   tunnel or peer failure and maintain NAT/NAPT contexts.  The L2TPv2
   HELLO message provides a simple, low-overhead method of doing this.

   The default values specified in [RFC2661] for L2TPv2 HELLO messages
   could result in a dead-end detection time of 83 seconds.  Although
   these retransmission timers and counters SHOULD be configurable (see
   Section 5.8 of [RFC2661]), these values may not be adapted for all
   situations, where a quicker dead-end detection is required, or where
   NAT/NAPT context needs to be refreshed more frequently.  In such
   cases, the SI/SC MAY use, in combination with L2TPv2 HELLO, LCP ECHO
   messages (Echo-Request and Echo-Reply codes) described in [RFC1661].
   When used, LCP ECHO messages SHOULD have a re-emission timer lower
   than the value for L2TPv2 HELLO messages.  The default value
   recommended in Section 6.5 of [RFC2661] for the HELLO message
   retransmission interval is 60 seconds.  When used, a set of suggested
   values (included here only for guidance) for the LCP ECHO message

   request interval is a default of 30 seconds, a minimum of 10 seconds,
   and a maximum of the lesser of the configured L2TPv2 HELLO
   retransmission interval and 60 seconds.

### 5.1.3. Tunnel Teardown

   Either the SI or SC can tear down the session and tunnel.  This is
   done as specified in Section 5.7 of [RFC2661], by sending a StopCCN
   control message.  There is no action specific to Softwires in this
   case.

### 5.1.4. Additional L2TPv2 Considerations

   In the Softwire "Hub and Spoke" framework, L2TPv2 is layered on top
   of UDP, as part of an IP-in-IP tunnel; Section 8.1 of [RFC2661]
   describes L2TP over UDP/IP.  Therefore, the UDP guidelines specified
   in [RFC5405] apply, as they pertain to the UDP tunneling scenarios
   carrying IP-based traffic.  Section 3.1.3 of [RFC5405] specifies that
   for this case, specific congestion control mechanisms for the tunnel
   are not necessary.  Additionally, Section 3.2 of [RFC5405] provides
   message size guidelines for the encapsulating (outer) datagrams,
   including the recommendation to implement Path MTU Discovery (PMTUD).

## 5.2. PPP Connection

   This section describes the PPP negotiations between the SI and SC in
   the Softwire context.

### 5.2.1. MTU

   The MTU of the PPP link presented to the SPH SHOULD be the link MTU
   minus the size of the IP, UDP, L2TPv2, and PPP headers together.  On
   an IPv4 link with an MTU equal to 1500 bytes, this could typically
   mean a PPP MTU of 1460 bytes.  When the link is managed by IPsec,
   this MTU SHOULD be lowered to take into account the ESP encapsulation
   (see [SW-SEC]).  The value for the MTU may also vary according to the
   size of the L2TP header, as defined by the leading bits of the L2TP
   message header (see [RFC2661]).  Additionally, see [RFC4623] for a
   detailed discussion of fragmentation issues.

### 5.2.2. LCP

   Once the L2TPv2 session is established, the SI and SC initiate the
   PPP connection by negotiating LCP as described in [RFC1661].  The
   Address-and-Control-Field-Compression configuration option (ACFC)
   [RFC1661] MAY be rejected.

5.2.3.  Authentication

   After completing LCP negotiation, the SI and SC MAY optionally
   perform authentication.  If authentication is chosen, Challenge
   Handshake Authentication Protocol (CHAP) [RFC1994] authentication
   MUST be supported by both the Softwire Initiator and Softwire
   Concentrator.  Other authentication methods such as Microsoft CHAP
   version 1 (MS-CHAPv1) [RFC2433] and Extensible Authentication
   Protocol (EAP) [RFC3748] MAY be supported.

   A detailed discussion of Softwire security is contained in [SW-SEC].

5.2.4.  IPCP

   The only Network Control Protocol (NCP) negotiated in the Softwire
   context is IPV6CP (see Section 5.2.4.1) for IPv6 as SPH, and IPCP
   (see Section 5.2.4.2) for IPv4 as SPH.

5.2.4.1.  IPV6CP

   In the IPv6-over-IPv4 scenarios (see Section 3.1), after the optional
   authentication phase, the Softwire Initiator MUST negotiate IPV6CP as
   defined in [RFC5072].  IPV6CP provides a way to negotiate a unique
   64-bit Interface-Identifier to be used for the address
   autoconfiguration at the local end of the link.

5.2.4.2.  IPv4CP

   In the IPv4-over-IPv6 scenarios (see Section 3.2), a Softwire
   Initiator MUST negotiate IPCP [RFC1332].  The SI uses IPCP to obtain
   an IPv4 address from the SC.  IPCP MAY also be used to obtain DNS
   information as described in [RFC1877].

5.3.  Global IPv6 Address Assignment to Endpoints

   In several scenarios defined in Section 3.1, global IPv6 addresses
   are expected to be allocated to Softwire endpoints (in addition to
   the Link-Local addresses autoconfigured using the IPV6CP negotiated
   interface identifier).  The Softwire Initiator assigns global IPv6
   addresses using the IPV6CP negotiated interface identifier and using
   Stateless Address Autoconfiguration [RFC4862], and/or using Privacy
   Extensions for Stateless Address Autoconfiguration [RFC4941], (as
   described in Section 5 of [RFC5072]), and/or using DHCPv6 [RFC3315].

   The Softwire Initiator of an IPv6 Softwire MUST send a Router
   Solicitation message to the Softwire Concentrator after IPV6CP is
   completed.  The Softwire Concentrator MUST answer with a Router
   Advertisement.  This message MUST contain the global IPv6 prefix of
   the PPP link if Neighbor Discovery is used to configure addresses of
   Softwire endpoints.

   If DHCPv6 is available for address delegation, the M bits of the
   Router Advertisement SHOULD be set.  The Softwire Initiator MUST then
   send a DHCPv6 Request to configure the address of the Softwire
   endpoint.

   Duplicate Address Detection ([RFC4861]) MUST be performed on the
   Softwire in both cases.

## 5.4. DHCP

   The Softwire Initiator MAY use DHCP to get additional information
   such as delegated prefix and DNS servers.

### 5.4.1. DHCPv6

   In the scenarios in Section 3.1, if the SI supports DHCPv6, it SHOULD
   send a Solicit message to verify if more information is available.

   If an SI establishing an IPv6 Softwire acts as a router (i.e., in the
   scenarios in Sections 3.1.2 and 3.1.4) it MUST include the Identity
   Association for Prefix Delegation (IA_PD) option [RFC3633] in the
   DHCPv6 Solicit message [RFC3315] in order to request an IPv6 prefix.

   When delegating an IPv6 prefix to the SI by returning a DHCPv6
   Advertise message with the IA_PD and IP_PD Prefix options [RFC3633],
   the SC SHOULD inject a route for this prefix in the IPv6 routing
   table in order to forward the traffic to the relevant Softwire.

   Configuration of DNS MUST be done as specified in [RFC3646] and
   transmitted according to [RFC3315] and [RFC3736].  In general, all
   DHCPv6 options MUST be transmitted according to [RFC3315] and
   [RFC3736].

### 5.4.2. DHCPv4

   An SI establishing an IPv4 Softwire MAY send a DHCP request
   containing the Subnet Allocation option [SUBNET-ALL].  This practice
   is not common, but it may be used to connect IPv4 subnets using
   Softwires, as defined in Sections 3.2.2 and 3.2.4.

   One Subnet-Request suboption MUST be configured with the 'h' bit set
   to '1', as the SI is expected to perform the DHCP server function.
   The 'i' bit of the Subnet-Request suboption SHOULD be set to '0' the
   first time a prefix is requested and to '1' on subsequent requests,
   if a prefix has been allocated.  The Prefix length suboption SHOULD
   be 0 by default.  If the SI is configured to support only specific
   prefix lengths, it SHOULD specify the longest (smallest) prefix
   length it supports.

   If the SI was previously assigned a prefix from that same SC, it
   SHOULD include the Subnet-Information suboption with the prefix it
   was previously assigned.  The 'c' and 's' bits of the suboption
   SHOULD be set to '0'.

   In the scenarios in Section 3.2, when delegating an IPv4 prefix to
   the SI, the SC SHOULD inject a route for this prefix in the IPv4
   routing table in order to forward the traffic to the relevant
   Softwire.

6.  Considerations about the Address Provisioning Model

   This section describes how a Softwire Concentrator may manage
   delegated addresses for Softwire endpoints and for subnets behind the
   Softwire Initiator.  One common practice is to aggregate endpoints'
   addresses and delegated prefixes into one prefix routed to the SC.
   The main benefit is to ease the routing scheme by isolating on the SC
   succeeding route injections (when delegating new prefixes for SI).

6.1.  Softwire Endpoints' Addresses

6.1.1.  IPv6

   A Softwire Concentrator should provide globally routable addresses to
   Softwire endpoints.  Other types of addresses such as Unique Local
   Addresses (ULAs) [RFC4193] may be used to address Softwire endpoints
   in a private network with no global connectivity.  A single /64
   should be assigned to the Softwire to address both Softwire
   endpoints.

   Global addresses or ULAs must be assigned to endpoints when the
   scenario "Host CPE as Softwire Initiator" (described in
   Section 3.1.1) is considered to be deployed.  For other scenarios,
   link-local addresses may also be used.

### 6.1.2. IPv4

A Softwire Concentrator may provide either globally routable or
private IPv4 addresses.  When using IPv4 private addresses [RFC1918]
on the endpoints, it is not recommended to delegate an IPv4 private
prefix to the SI, as it can lead to a nested-NAT situation.

The endpoints of the PPP link use host addresses (i.e., /32),
negotiated using IPCP.

## 6.2. Delegated Prefixes

### 6.2.1. IPv6 Prefixes

Delegated IPv6 prefixes should be of global scope if the IPv6
addresses assigned to endpoints are global.  Using ULAs is not
recommended when the subnet is connected to the global IPv6 Internet.
When using IPv6 ULAs on the endpoints, the delegated IPv6 prefix may
be either of global or ULA scope.

Delegated IPv6 prefixes are between /48 and /64 in length.  When an
SI receives a prefix shorter than 64, it can assign different /64
prefixes to each of its interfaces.  An SI receiving a single /64 is
expected to perform bridging if more than one interface is available
(e.g., wired and wireless).

### 6.2.2. IPv4 Prefixes

Delegated IPv4 prefixes should be routable within the address space
used by assigned IPv4 addresses.  Delegate non-routable IPv4 prefixes
(i.e., private IPv4 prefix over public IPv4 addresses or another
class of private IPv4 addresses) is not recommended as a practice for
provisioning and address translation should be considered in these
cases.  The prefix length is between /8 and /30.

## 6.3. Possible Address Provisioning Scenarios

This section summarizes the different scenarios for address
provisioning with the considerations given in the previous sections.

6.3.1.  Scenarios for IPv6

   This table describes the possible combination of IPv6 address scope
   for endpoints and delegated prefixes.

   | Endpoint IPv6 Address | Delegated Global IPv6 Prefix | Delegated ULA IPv6 Prefix |
   |---|---|---|
   | Link Local | Possible | Possible |
   | ULA | Possible | Possible |
   | Global | Possible | Possible, but Not Recommended |

                       Table 1: Scenarios for IPv6

6.3.2.  Scenarios for IPv4

   This table describes the possible combination of IPv4 address scope
   for endpoints and delegated prefixes.

   | Endpoint IPv4 Address | Delegated Public IPv4 Prefix | Delegated Private IPv4 Prefix |
   |---|---|---|
   | Private IPv4 | Possible | Possible, but Not Recommended when using NAT (cf. Section 6.1.2) |
   | Public IPv4 | Possible | Possible, but NAT usage is recommended (cf. Section 6.2.2) |

                       Table 2: Scenarios for IPv4

7.  Considerations about Address Stability

   A Softwire can provide stable addresses even if the underlying
   addressing scheme changes, by opposition to automatic tunneling.  A
   Softwire Concentrator should always provide the same address and
   prefix to a reconnecting user.  However, if the goal of the Softwire
   service is to provide a temporary address for a roaming user, it may
   be provisioned to provide only a temporary address.

   The address and prefix are expected to change when reconnecting to a
   different Softwire Concentrator.  However, an organization providing
   a Softwire service may provide the same address and prefix across
   different Softwire Concentrators at the cost of a more fragmented
   routing table.  The routing fragmentation issue may be limited if the
   prefixes are aggregated in a location topologically close to the SC.
   This would be the case, for example, if several SCs are put in
   parallel for load-balancing purpose.

8.  Considerations about RADIUS Integration

   The Softwire Concentrator is expected to act as a client to a AAA
   server, for example, a RADIUS server.  During the PPP authentication
   phase, the RADIUS server may return additional information in the
   form of attributes in the Access-Accept message.

   The Softwire Concentrator may include the Tunnel-Type and Tunnel-
   Medium-Type attributes [RFC2868] in the Access-Request messages to
   provide a hint of the type of Softwire being configured.

8.1.  Softwire Endpoints

8.1.1.  IPv6 Softwires

   If the RADIUS server includes a Framed-Interface-Id attribute
   [RFC3162], the Softwire Concentrator must send it to the Softwire
   Initiator in the Interface-Identifier field of its IPV6CP
   Configuration Request message.

   If the Framed-IPv6-Prefix attribute [RFC3162] is included, that
   prefix must be used in the router advertisements sent to the SI.  If
   Framed-IPv6-Prefix is not present but Framed-IPv6-Pool is, the SC
   must choose a prefix from that pool to send RAs.

8.1.2.  IPv4 Softwires

   If the Framed-IP-Address attribute [RFC2865] is present, the Softwire
   Concentrator must provide that address to the Softwire Initiator
   during IPCP address negotiation.  That is, when the Softwire
   Initiator requests an IP address from the Softwire Concentrator, the
   address provided should be the Framed-IP-Address.

8.2.  Delegated Prefixes

8.2.1.  IPv6 Prefixes

   If the attribute Delegated-IPv6-Prefix [RFC4818] is present in the
   RADIUS Access-Accept message, it must be used by the Softwire
   Concentrator for the delegation of the IPv6 prefix.  Since the prefix
   delegation is performed by DHCPv6 and the attribute is linked to a
   username, the SC must associate the DHCP Unique Identifier (DUID) of
   a DHCPv6 request to the tunnel it came from and its user.

   Interaction between RADIUS, PPP, and DHCPv6 server may follow the
   mechanism proposed in [RELAY-RAD].  In this case, during the Softwire
   authentication phase, PPP collects the RADIUS attributes for the user
   such as Delegated-IPv6-Prefix.  A specific DHCPv6 relay is assigned
   to the Softwire.  The DHCPv6 relay fills in these attributes in the
   Relay agent RADIUS Attribute Option (RRAO) DHCPv6 option, before
   forwarding the DHCPv6 requests to the DHCPv6 server.

8.2.2.  IPv4 Prefixes

   RADIUS does not define an attribute for the delegated IPv4 Prefix.
   Attributes indicating an IPv4 prefix and its length (for instance the
   combination of the Framed-IP-Address and Framed-IP-Netmask attributes
   [RFC2865]) may be used by the Softwire Concentrator to delegate an
   IPv4 prefix to the Softwire Initiator.  The Softwire Concentrator
   must add a corresponding route with the Softwire Initiator as next-
   hop.

   As this practice had been used, the inclusion of the Framed-IP-
   Netmask attribute along with the Framed-IP-Address attribute tells
   the Softwire Concentrator to delegate an IPv4 prefix to the Softwire
   Initiator (e.g., in the IPv4-over-IPv6 scenarios where the Softwire
   Initiator is a router, see Sections 3.2.2 and 3.2.4), as the SC
   should forward packets destined to any IPv4 address in the prefix to
   the SI.

9.  Considerations for Maintenance and Statistics

   Existing protocol mechanics for conveying adjunct or accessory
   information for logging purposes, including L2TPv2 and RADIUS
   methods, can include informational text that the behavior is
   according to the Softwire "Hub and Spoke" framework (following the
   implementation details specified in this document).

9.1. RADIUS Accounting

   RADIUS Accounting for L2TP and PPP are documented (see Section 4.3).

   When deploying Softwire solutions, operators may experience
   difficulties to differentiate the address family of the traffic
   reported in accounting information from RADIUS.  This problem and
   some potential solutions are described in [SW-ACCT].

9.2. MIBs

   MIB support for L2TPv2 and PPP are documented (see Section 4.4).
   Also, see [RFC4293].

10. Security Considerations

   One design goal of the "Hub and Spoke" problem is to very strongly
   consider the reuse of already deployed protocols (see [RFC4925]).
   Another design goal is a solution with very high scaling properties.
   L2TPv2 [RFC2661] is the phase 1 protocol used in the Softwire "Hub
   and Spoke" solution space, and the L2TPv2 security considerations
   apply to this document (see Section 9 of [RFC2661]).

   The L2TPv2 Softwire solution adds the following considerations:

   o  L2TP Tunnel Authentication (see Sections 5.1.1 and 9.1 of
      [RFC2661]) provides authentication at tunnel setup.  It may be
      used to limit DoS attacks by authenticating the tunnel before L2TP
      and PPP resources are allocated.

   o  In a Softwire environment, L2TPv2 AVPs do not transport sensitive
      data, and thus the L2TPv2 AVP hiding mechanism is not used (see
      Section 5.1.1).

   o  PPP CHAP [RFC1994] provides basic user authentication.  Other
      authentication protocols may additionally be supported (see
      Section 5.2.3).

   L2TPv2 can also be secured with IPsec to provide privacy, integrity,
   and replay protection.  Currently, there are two different solutions
   for security L2TPv2 with IPsec:

   o  Securing L2TPv2 using IPsec "version 2" (IKEv1) is specified in
      [RFC3193], [RFC3947], and [RFC3948].  When L2TPv2 is used in the
      Softwire context, the voluntary tunneling model applies.
      [RFC3193] describes the interaction between IPsec and L2TPv2, and

      is deployed.  [RFC3193] MUST be supported, given that deployed
      technology must be very strongly considered [RFC4925] for this
      'time-to-market' solution.

   o  [SW-SEC] also specifies a new (incompatible) solution for securing
      L2TPv2 with IPsec "version 3" (IKEv2).  Section 3.5 of [SW-SEC]
      describes the advantages of using IKEv2, and this solution needs
      to be considered for future phases.

   Additional discussion of Softwire security is contained in [SW-SEC].

11.  Acknowledgements

   The authors would like to acknowledge the following contributors who
   provided helpful input on this document: Florent Parent, Jordi Palet
   Martinez, Ole Troan, Shin Miyakawa, Carl Williams, Mark Townsley,
   Francis Dupont, Ralph Droms, Hemant Singh, and Alain Durand.

   The authors would also like to acknowledge the participants in the
   Softwires interim meetings held in Hong Kong, China, and Barcelona,
   Spain.  The minutes for the interim meeting at the China University -
   Hong Kong (February 23-24, 2006) are at
   <http://www.ietf.org/proceedings/06mar/isoftwire.html>.  The minutes
   for the interim meeting at Polytechnic University of Catalonia -
   Barcelona (September 14-15, 2006) are reachable at
   <http://www.ietf.org/proceedings/06nov/isoftwire.html>.  The
   Softwires auxiliary page at <http://bgp.nu/~dward/softwires/>
   contains additional meeting information.

   During and after the IETF Last Call, useful comments and discussion
   were provided by Jari Arkko, David Black, Lars Eggert, Pasi Eronen,
   and Dan Romascanu.

## 12. References

### 12.1. Normative References

   

12.2.  Informative References

   

Authors' Addresses

   Bill Storer
   Cisco Systems
   170 W Tasman Dr
   San Jose, CA  95134
   USA

   EMail: bstorer@cisco.com

   Carlos Pignataro (editor)
   Cisco Systems
   7200 Kit Creek Road
   PO Box 14987
   Research Triangle Park, NC  27709
   USA

   EMail: cpignata@cisco.com

   Maria Alice Dos Santos
   Cisco Systems
   170 W Tasman Dr
   San Jose, CA  95134
   USA

   EMail: mariados@cisco.com

   Bruno Stevant (editor)
   TELECOM Bretagne
   2 rue de la Chataigneraie CS17607
   Cesson Sevigne,    35576
   France

   EMail: bruno.stevant@telecom-bretagne.eu

   Laurent Toutain
   TELECOM Bretagne
   2 rue de la Chataigneraie CS17607
   Cesson Sevigne,    35576
   France

   EMail: laurent.toutain@telecom-bretagne.eu

   Jean-Francois Tremblay
   Videotron Ltd.
   612 Saint-Jacques
   Montreal, QC  H3C 4M8
   Canada

   EMail: jf@jftremblay.com